\documentclass[conference]{IEEEtran}
\IEEEoverridecommandlockouts
\usepackage{cite}
\usepackage{amsmath,amssymb,amsfonts}

\usepackage{graphicx}
\usepackage{textcomp}
\usepackage{xcolor}
\def\BibTeX{{\rm B\kern-.05em{\sc i\kern-.025em b}\kern-.08em
    T\kern-.1667em\lower.7ex\hbox{E}\kern-.125emX}}
\usepackage{float}
\graphicspath{ {./images/} }
\usepackage{enumerate}
\usepackage{comment}
\usepackage[hyphens]{url}
\usepackage{soul}
\usepackage[english]{babel}
\usepackage{xspace}
\usepackage{siunitx}
\usepackage[numbers]{natbib}
\usepackage{caption}
\usepackage{algorithm}
\usepackage{algpseudocode}
\usepackage{pifont}
\usepackage{graphicx}
\usepackage{framed}
\usepackage{multirow}
\usepackage{stmaryrd}  
\usepackage{subfigure}

\DeclareCaptionType[fileext=ext]{Protocol}[Protocol]
\usepackage{tabularx}

\newcommand{\Sys}{{AegisBlock}\xspace}

\newcommand{\patientSKPK}{($x^b_p$, $y^b_p$)}
\newcommand{\patientIDKPair}{$({x}^{\text{ID}}_p$, ${y}^{\text{ID}}_p)$}

\newcommand{\patientBPK}{$y^b_p$}

\newcommand{\hospitalBPK}{$y^b_h$}

\newcommand{\hospitalSKPK}{($x^b_h$, $y^b_h$)}
\newcommand{\hospitalIDKPair}{$({x}^{\text{ID}}_h$, ${y}^{\text{ID}}_h)$}

\newcommand{\patientPKG}{${y}^{\text{}}_{p}$}

\newcommand{\hospitalPKG}{$y_h$}

\newcommand{\symK}{$s^b_p$}
\newcommand{\hospitalGroupK}{$y_H$}
\graphicspath{{figures/}} 

\makeatletter
\newcommand{\linebreakand}{%
  \end{@IEEEauthorhalign}
  \hfill\mbox{}\par
  \mbox{}\hfill\begin{@IEEEauthorhalign}
}
\makeatother

\begin{document}


\title{\Sys: A Privacy-Preserving Medical Research Framework using Blockchain}










\author{
\IEEEauthorblockN{Calkin Garg*\thanks{*These authors contributed equally.}}
\IEEEauthorblockA{\textit{Dept. of Computer Science} \\
\textit{Georgia Institute of Technology}\\
cgarg35@gatech.edu}
\and
\IEEEauthorblockN{Omar Rios Cruz*}
\IEEEauthorblockA{\textit{Dept. of Computer Science} \\
\textit{California State University, Stanislaus}\\
orioscruz@csustan.edu}
\and
\IEEEauthorblockN{Tessa E. Andersen}
\IEEEauthorblockA{\textit{Dept. of Computer Science} \\
\textit{Brigham Young University}\\
tessa343@byu.edu}

\linebreakand
\IEEEauthorblockN{Gaby G. Dagher} 
\IEEEauthorblockA{\textit{Dept. of Computer Science} \\
\textit{Boise State University}\\
gabydagher@boisestate.edu}
\and
\IEEEauthorblockN{Donald Winiecki} 
\IEEEauthorblockA{\textit{Dept. of OPWL} \\
\textit{Boise State University}\\
dwiniecki@boisestate.edu}


\and
\IEEEauthorblockN{Min Long}
\IEEEauthorblockA{\textit{Dept. of Computer Science}\\
\textit{Boise State University}\\
minlong@boisestate.edu}
}

\maketitle 
\begin{abstract}




Due to HIPAA and other privacy regulations, it is imperative to maintain patient privacy while conducting research on patient health records. In this paper, we propose \Sys, a patient-centric access controlled framework to share medical records with researchers such that the anonymity of the patient is maintained while ensuring the trustworthiness of the data provided to researchers. \Sys allows for patients to provide access to their medical data, verified by miners. A researcher submits a time-based range query to request access to records from a certain patient, and upon patient approval, access will be granted. Our experimental evaluation results show that \Sys is scalable with respect to the number of patients and hospitals in the system, and efficient with up to 50\% of malicious miners.




\end{abstract}



\section{\uppercase{Introduction}}\label{sec:introduction}
Across all aspects of research, there are often complications related to the access of data. This is especially the case when the data comes from "human subjects"~\cite{humanSubjects}. Protection of privacy related to personal health information (PHI) is not only ethically virtuous and a common good when accomplished for all people; it is also mandated in the USA with HIPAA law~\cite{HIPAASummary}. HIPAA was explicitly intended as a strong precedent for handling patient privacy and security to prevent mismanagement of data. HIPAA provides patients with the confidence that their PHI can only be used and shared with individuals who must have it to assist with their medical care.



HIPAA also allows patients to expand or limit access to components of their PHI to groups such as researchers who might find those data valuable in aggregate~\cite{researchersHIPAA}. This means that even if researchers obtain permission to access data by complying with HIPAA regulations, the representativeness of data for demographic groups may be reduced as a result of selection bias when patients decline to share their PHI with researchers~\cite{selectionBias}. We focus on observational researchers conducting cohort studies that discern the change of multiple patients with similar characteristics over a set period of time ranging from a few years to decades~\cite{cohortStudyTime}. 

These aforementioned HIPAA guidelines point to a dilemma for medical researchers, who must maintain the ethical necessity of these protections while still seeking large sets of data that are representative of populations of interest~\cite{researchChallenges1}. This dilemma impacts the utility of data for researchers and may also have an adverse impact on the validity of research, which could have a follow-up negative influence on the value of applications of their research~\cite{researchChallenges}.

A clear solution to this problem would allow data from a patient's personal health record (PHR) to be available in an environment that maintains HIPAA regulations and follows ethical values in a safe and secure manner. For our purposes, we consider "safe and secure" to mean a system where only patients, hospitals, and researchers are allowed access to only portions of the data they require for their specific purposes. In order to preserve patient privacy through anonymity, we will require multiple methods of encryption and verification while also masking the identities of individuals. However, this system would only work when patients have previously agreed to share their data with researchers in addition to the necessary medical providers. One technology that can assist with the delivery of data while also providing measures to make sure only authorized individuals have access is a Permissioned Blockchain~\cite{blockchainDataSharing}. Blockchain has emerged as a viable solution for managing access to patient data. For one, the system requires consensus before data can be shared~\cite{blockchainConsensus}. Blockchain is also an immutable and decentralized technology that incorporates various protocols to facilitate transactions~\cite{blockchainDecentralized}, which creates a safe system in which patients can input their data and researchers can request it with the consent of the appropriate patient. 

In this paper, we utilize blockchain technology to introduce \Sys, patient-focused system for sharing data with researchers. \Sys behaves as an \textit{access control system}, allowing patients to have full authority to decide who has access to their data, while also allowing researchers to submit time-based range queries to support observational cohort studies. \Sys is a permissioned blockchain, where we assume all users in our system are a valid member of their respective party, validated by a trusted third party. We rely on the immutability property of blockchain to allow researchers to verify the integrity and accuracy of the information they receive. 

\subsection{Contribution}
The contributions of this paper are as follows:

\begin{enumerate}

    \item We propose \Sys, a proof-of-concept patient-centric permissioned blockchain framework which serves as an access mechanism for sharing medical data with researchers.

     \item We incorporate methods to reference patients' previous blocks while still protecting patients' privacy.

     \item \Sys facilitates researchers' requests for patients' personal health records to conduct cohort studies, while ensuring the trustworthiness of the shared data.

     \item We conducted experiment evaluation on \Sys. The results demonstrate its scalability with respect to the number of users (patients, hospitals, and researchers) in the system. They also show that \Sys is robust against up to 50\% of malicious miners (hospitals). 
     
\end{enumerate}

\section{Related Work}\label{sec:related_work}
Prior research has attempted to address the delivery of data from a patient to a vast multitude of parties. However, a majority of these papers only establish a "group" as the end user that collects data from patients. 

\subsection{Patient-Centric Blockchain System}
~\citeauthor{omniPHR} proposed a system titled OmniPHR~\cite{omniPHR}, which handles PHR with respect to data maintenance. Similarly to OmniPHR,~\citeauthor{action-EHR} also proposed their own system, ACTION-EHR~\cite{action-EHR}. What separates ACTION-EHR from their related work is the use of smart contracts through Hyperledger Fabric. ACTION-EHR also mentions the use PBFT as a consensus mechanism used within permissioned blockchains. In both works, patients benefit from being in their systems by having all their records stored within a secure and private blockchain. Hospitals benefit while in the system because they can collect the same patient's data across multiple visits at different locations, allowing them to perform better evaluations when prescribing care. However, the sharing of patient data among differing medical institutions does raise concerns about anonymity, especially regarding access to patient data from hospitals the patient has never been to. \Sys avoids this problem by having the patient choose who their data goes to.

~\citeauthor{medRec} proposed MedRec~\cite{medRec}, a proof-of-work patient-centric blockchain system in which stakeholders can query through a string that lists specific patient data. In MedRec, smart contracts are utilized through medical records and their associated viewing rights. The main purpose of MedRec was to create a system where patients are able to maintain their own data, such as OmniPHR.

~\citeauthor{EHRZKP} proposed a system that revolves around the use of ZKPs to secure patient anonymity~\cite{EHRZKP}. Our system relies on such technologies; however, their usage is merely meant to create a new account on a web application. \Sys uses ZKPs to validate patients and hospitals without having to reveal their identities when their blocks are being mined. This way, any block that has already been created can be trusted that all the information inside is valid.

\begin{table*}[h]

\begin{center}

\caption{Comparative evaluation of main features in closely related work including our proposed work \Sys.}

\label{relatedWorksTable}

\begin{tabular}{|l|c|c|c|c|c|c|}
\hline
 
\multicolumn{1}{|l|}{\textbf{Papers}} & \multicolumn{2}{c}{\textbf{Protection of Patient}} & \multicolumn{1}{|c|}{\textbf{Researcher Focused}} & \multicolumn{1}{c|}{\textbf{Smart Contracts}} & \multicolumn{1}{c|}{\textbf{Consensus Mechanism}}\\
 
\cline{2-3} &\textbf{Data} & \textbf{Anonymity} & & &\\ 
\hline
 
\citet{omniPHR} & \checkmark &  & & & \\
\hline

\citet{action-EHR} & \checkmark & \checkmark & \checkmark & \checkmark & PBFT\\
\hline

\citet{EHRZKP} & \checkmark & \checkmark &  &  & \\
\hline
 
\citet{healthChain} & \checkmark & \checkmark &  & \checkmark & PoC\\
\hline

\citet{homomorphicEncryption} & \checkmark &  & \checkmark &  & PoC\\
\hline

\citet{hierarchicalBlockchain} & \checkmark &  &  & \checkmark & \\
\hline

\citet{medRec} & \checkmark & & & \checkmark & PoW\\
\hline

\hline
This Paper: \Sys & \checkmark & \checkmark & \checkmark &  & PoC\\
\hline
 
\end{tabular}

\end{center}

\end{table*}

~\citeauthor{healthChain} describe a proof-of-concept, patient-centric, blockchain system in which patients, through the use of smart contracts, can share their data~\cite{healthChain}. HealthChain utilizes proxy re-encryption, which is a tool we do not need to use, which simplifies our process.

Other papers such as~\cite{Ancile},~\cite{incrementalUpdates}, and~\cite{searchableBlockchain} are all for the sake of maintaining and preserving anonymity with Patient Health Records (PHR). These papers differ with complex smart contract usage, incremental updates, and the ability to query through the blockchain.

\subsection{Patient-Centric Blockchain System with a Focus on Researchers}
~\citeauthor{homomorphicEncryption} created a proof-of-concept, patient-centric blockchain design that has researchers as the end user~\cite{homomorphicEncryption}. However, they use computationally expensive mechanisms such as Fully Homomorphic Encryption, while also leaving out specific information such as the mining process or key generation. Our work focuses on offering these protocols within a patient-centric, researcher-focused blockchain.

 ~\citeauthor{frameworkApp} also created a framework that focuses on researchers, but not as the primary end user~\cite{frameworkApp}. Their implementation of an application along with patients deciding how they can deal with researcher requests does make it similar to our paper. However, after a researcher has been granted access to view a PHR, they will be required to pay. This raises multiple ethical concerns, which we are not comfortable having within our system since we would have to explore the ethics of selling personal medical data, even if the patient provided consent. 

~\citeauthor{hierarchicalBlockchain} had proposed a management system to allow a patient to pick who their information goes to~\cite{hierarchicalBlockchain}, however, the systems do not mention anything about the preservation of anonymity. This system incorporates the smart contracts, split between staff member registration and access verification. \Sys preserves anonymity through every step of the blockchain. Furthermore,~\cite{hierarchicalBlockchain} does not mention how a block is created, what is stored in the block, or who the miners are, all of which are discussed in \Sys. Table~\ref{relatedWorksTable} provides a comparative evaluation of closely related works.

\section{Problem Formulation}\label{sec:problem_formulation}
We identified many problems within the current world of clinical research, which we attempt to solve with \Sys. 

First, a researcher would want to identify the group that they want to study. This selection process is usually performed with certain characteristics in mind~\cite{typesOfCohortStudy}, be it diseases, infections, etc. After identifying a specific group, researchers would usually have to go through a hospital. After gathering the consent of the participants, the researcher would proceed to conduct their studies.

Before a researcher attempts to conduct a cohort study, they would first need to go through a hospital's governing body. Researchers would need to obtain approval from an Institutional Review Board, then they would need consent from patients~\cite{researcherObstacles}. From this point, the researcher would collect as much data as they need. One of the problems with observational studies is the amount of time it takes. This problem can take up to several years at a time~\cite{researcherTime}. By allowing a researcher to access data from any range in time, we seek to cut down on the amount of time it takes to conduct research. As we are focused on improving the process for observational studies, we will not discuss the requirement of clinical trials in this paper.

From the perspective of a patient, \Sys focuses on giving them full authority when it comes to handling their data. One problem with a cohort study is that one needs to inherently trust that the hospital or other covered entity handling personal data takes the appropriate measures to ensure that it has gone through a sufficient amount of differential privacy or has been de-identified entirely~\cite{dataDeidentification}. Another problem patients face is what data they can choose to give away. \Sys creates a system in which researchers request data from a patient without accessing the patient's entire health record, similar to how actual studies are conducted. With other studies, there is an agreement to allow active monitoring over a set period of time. \Sys creates a system which allows a researcher to access data from a certain time frame, without the need for active monitoring.

\subsection{Notation}
Table~\ref{tbl:notations} shows all the symbols we will use for the rest of the paper.

\begin{table}[h]
\centering
\small
\renewcommand{\arraystretch}{1} 
\begin{tabular}{ |r |l|  }
\hline

Symbol & Description \\
\hline

\textit{p} & Patient\\
\textit{r} & Researcher\\
\textit{h} & Hospital\\
\textit{b} & Block ID\\
\textit{$n^b$} & Nonce\\

\hline
\patientIDKPair & Patient's identity key pair\\
\patientSKPK & Patient's block key pair\\
\symK & Patient-block-specific symmetric key\\
\hospitalIDKPair & Hospital's identity key pair\\
\hospitalSKPK & Hospital's block key pair\\
\hospitalGroupK & Hospitals' Group Public Key\\
($x^{\text{ID}}_r, y^{\text{ID}}_r$)  & Researcher Identity Key Pair\\
\hline

\end{tabular}
\caption{Notations} \label{tbl:notations}
\end{table}


\subsection{Adversarial Model}
Within \Sys, there are three types of parties that interact with our system. These are \textit{patients}, \textit{hospitals}, and \textit{researchers}. A patient, along with their hospital, collaborates to upload a specific PHR information for a patient onto a block on \Sys. The hospitals will also be part of the mining group that validates whether or not a block should be uploaded onto \Sys. Since \Sys is a permissioned blockchain, a trusted third party is utilized to authenticate patients, hospitals, and researchers. We describe below the possible malicious behavior of each party in the system. 


\noindent \textbf{Patients.} A malicious patient may attempt to sign or upload a block with their set of keys while being an invalid user. 

\noindent \textbf{Hospitals.} A malicious hospital may attempt to sign a block with their set of keys while being an invalid user. Additionally, while a hospital is participating in the consensus protocol of a block, they may immediately attempt to decline signing a block, even though it is valid. A malicious hospital may also attempt to upload a patient block without the consent of the patient.

\noindent \textbf{Researchers.} A researcher may attempt to use the information the patient provides to access blocks outside the allowed range, or attempt to identify which specific blocks they are receiving data from. 

We assume that the number of malicious miners (hospitals) never exceeds 50\%. We also assume that there is no collusion between parties in the system.






\section{Solution: \Sys}


\begin{figure*}[t!]

    \centering
    \includegraphics[width = 1 \linewidth]{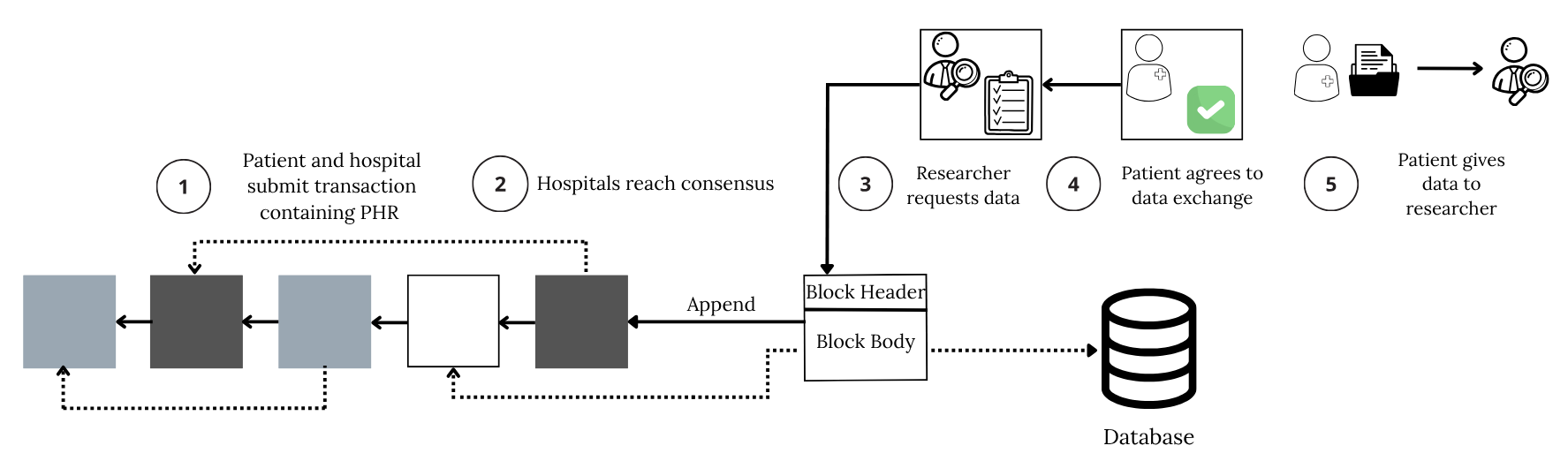}
    \caption{\Sys Overview. Each set of blocks with the same color represents different medical records for the same patient.}
    \label{blockchainOverview}
    
\end{figure*}

Our proposed system offers a novel framework for access control of medical data between researchers and patients. \Sys incorporates a private, permissioned blockchain with ZKPs to ensure patient privacy while giving researchers access to data they need for cohort studies. Within \Sys, we include two protocols, with one focused on patients sharing their data on the blockchain and the other allowing researchers access to patient data.

\Sys utilizes a private, permissioned blockchain, which means that patients and hospitals will be registered through a third party that verifies their identities before they are allowed to access \Sys. This increases security as only verified users can add medical data to \Sys, therefore decreasing malicious attacks where fake medical data is added. In addition to the verification of patients and hospitals, the researchers' identities are also verified before being added to the blockchain. This reduces the chance for malicious users to request access to medical data and increases trust between the researcher and the patients within \Sys.

The first protocol is for patient block creation. Patients will submit their PHR onto the chain, which is verified by their corresponding hospital. Other hospitals will act as miners on the blockchain and reach consensus on the data based on the validity of the data, the patient, and the hospital. Our second protocol is used to allow researchers to request and access PHR on the chain for their research. This protocol starts with a researcher requesting access to data within a specified range. The patient will then sign off on this request and give the data to the researcher. Our \Sys framework can be seen in Figure \ref{blockchainOverview}.

\subsection{Phase I: Patient Health Record (PHR) Submission}
The first phase of \Sys includes a patient submitting their PHR to be incorporated into the blockchain. A patient and their corresponding hospital will work together to create a block. The block will contain a header and body with differing information. Before submitting a block, the patient will generate a new private/public key pair for the block they are submitting to, \patientSKPK, along with a new symmetric key, \symK, for the same block. The hospital will also create a new private/public key pair, \hospitalSKPK, unique to this block. The patient will then encrypt their data using the symmetric key and store it in an off-chain database. 

\subsubsection{PHR Block Creation}
\begin{figure}[h!]

    \centering
    \includegraphics[width=0.5\linewidth]{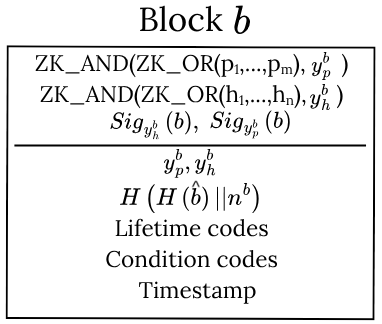}
    \caption{Patient Block Structure}
    \label{fig:transactionOverview}
    
\end{figure}

Within the PHR Block Creation section, we allow patients and hospitals to submit data on the blockchain for researcher access. After the patient and the related hospital have registered in \Sys, the patient can agree to submit their PHR to be incorporated into the blockchain. In this process, both the patient and the hospital will submit data to verify their identity and the integrity of the data. As seen in Figure \ref{fig:transactionOverview}, the header will contain information that is used to verify the authenticity of both parties. The body of the block will contain information that is primarily relevant to a PHR. 

\begin{Protocol*}[h]

    \begin{framed}
    
    \textbf{Block Creation Protocol}
    
    \footnotesize
    
    \textbf{Input: \:\: Patient data recorded this visit: $D_b$, hash of Patient's most recent block: $H(\hat{b}-1)$} 
    
    \textbf{Output: \:\:New block, $b$}
    
    \hrulefill

    The patient, $p$, constructs a new block for their data $D_b$, signed by both them and the hospital, $h$, such that access to encrypted $D_b$ can be granted to interested researchers.    

        \begin{enumerate}            






        \item $p$ creates a new block specific public key, \patientBPK.
        \label{p_new_key}

        \item $h$ creates a new block specific public key, \hospitalBPK.
        \label{h_new_key}

        \item $p$ submits a ZK\_AND(ZK\_OR($p_1$, \dots, $p_m$), \patientBPK)
        \label{p_zkp}

        \item $h$ submits a ZK\_AND(ZK\_OR($h_1$, \dots, $h_n$), \hospitalBPK)
        \label{h_zkp}

        \item $p$ submits their lifetime and visit specific condition codes.
        \label{codes}
  
        \item $p$ submits a hash to the block: $H(H(\hat{b})||n^b)$, where $n^b$ is a 256 bit nonce specific to $b$.
        \label{hash_on_block}

            \begin{enumerate}

                \item $H(\hat{b})$ is equal to: $H($\symK $\, || \, Ptr(D_b) \, || \, H(D_b) \, || \, H(\hat{b}-1) \,)$.
                \label{hat_definition}
                    
            \end{enumerate}

        \item $p$ uploads their public key to the block: \patientBPK.
        \label{p_upload_pk}

        \item $h$ uploads their public key to the block: \hospitalBPK.
        \label{h_upload_pk}

        \item If all information is correct, $p$ will sign off on the block: $Sig_{y^b_p}$
        \label{p_sign_block}

        \item If all information is correct, $h$ will sign off on the block: $Sig_{y^b_h}$
        \label{h_sign_block}

        \item $p$ submits new block, $b$, to be confirmed. 
        \label{return_new_block}
                    
        \end{enumerate}
        
    \normalsize	
    
    \end{framed}
    
    \caption{Block Creation Protocol}
    
    \label{Protocol:blockCreationProtocol}
 
\end{Protocol*}

\paragraph{Block Body}

Protocol~\ref{Protocol:blockCreationProtocol} details the process for creating a new block. In Steps~\ref{p_new_key}-\ref{h_new_key}, both the patient and hospital create new public keys. In Step~\ref{codes}, the patient will submit a sequence of bits that reveal their lifetime and visit specific conditions. A 0 in a specific position indicates that they do not have that condition, and a 1 indicates they do have the condition. The patient submits the bits in plaintext so that researchers can scan through all the blocks on the blockchain and only fork those of interest to them.


Next, in Step~\ref{hash_on_block}, the patient will submit a hash, $H(H(\hat{b})||n^b)$, where $H(\hat{b}) = H($\symK $\, || \, Ptr(D_b) \, || \, H(D_b) \, || \, H(\hat{b}-1) \,)$. We have the patient submit this hash to prevent researchers from looking beyond within their record range. The inclusion of the nonce, $n^b$, ensures that it is only necessary to reveal the very last block within a contiguous range of blocks while still allowing the researcher to verify that nothing in the range was tampered with. It also prevents against malicious researchers reconstructing these hashes and attempting to go through \Sys to find a match, which would inform them as to where the data they are looking at is coming from. If they are able to find these blocks, they could learn patient aliases and other information that is not required for what they want to do.

Then, in Step~\ref{p_upload_pk}-~\ref{h_upload_pk}, the patient and hospital will submit their newly generated public keys, \patientBPK, and \hospitalBPK. We do this so the miners can use it to verify the ZKP in the block header for the patient.


\paragraph{Block Header}

Contained inside of the header are non-interactive ZKP transcripts of both the hospitals and patients. In Steps~\ref{p_zkp}-\ref{h_zkp}, both the patient and hospital submit ZKPs to prove that these public keys are owned by a valid patient and a valid hospital, respectively. The patient will run a ZKP on their block-specific public key, \patientBPK. This shows the miners that whoever submitted the block knows the associated private key. The patient will also run a ZK-OR on all the public keys within the patient registry. This shows the miners that the patient knows a private key corresponding to a public key that belongs to a valid patient in the registry. Finally, the patient will run a ZK-AND where they provide their ZKP for this new public key and their ZK-OR as inputs. This combination proves to the miners that whoever controls this new public key is also a valid patient within the registry. We repeat the same steps for the hospital to ensure the hospital creating this block is valid. This combination is used by the miners to verify the identities of the patient and hospital without revealing who they each are. By honoring the privacy of the patient and the hospital, we remain ethically virtuous as required by HIPAA guidelines, while also providing a service to researchers.

After the ZKPs have been submitted, we proceed to Steps~\ref{p_sign_block} and~\ref{h_sign_block} where both the patient and the hospital sign off on the block, represented as $Sig_{y^b_p}(b)$ and $Sig_{y^b_h}(b)$, respectively. Both signatures are created using the new public key generated for the patient and the hospital. When the patient submits the block, they will sign it to verify that the data they submitted is valid. The associated hospital signs the block to verify that the data is accurate and has not been tampered with. When reaching consensus, miners can use these signatures to verify that the data is accurate and has not been tampered with, while also checking if the public keys associated with the signature exist and have been verified via the ZKP. Finally, in Step~\ref{return_new_block}, new block $b$ is submitted to the miners (hospitals) for confirmation.\\

\begin{figure*}[h]

    \centering
    \includegraphics[width = .80 \linewidth]{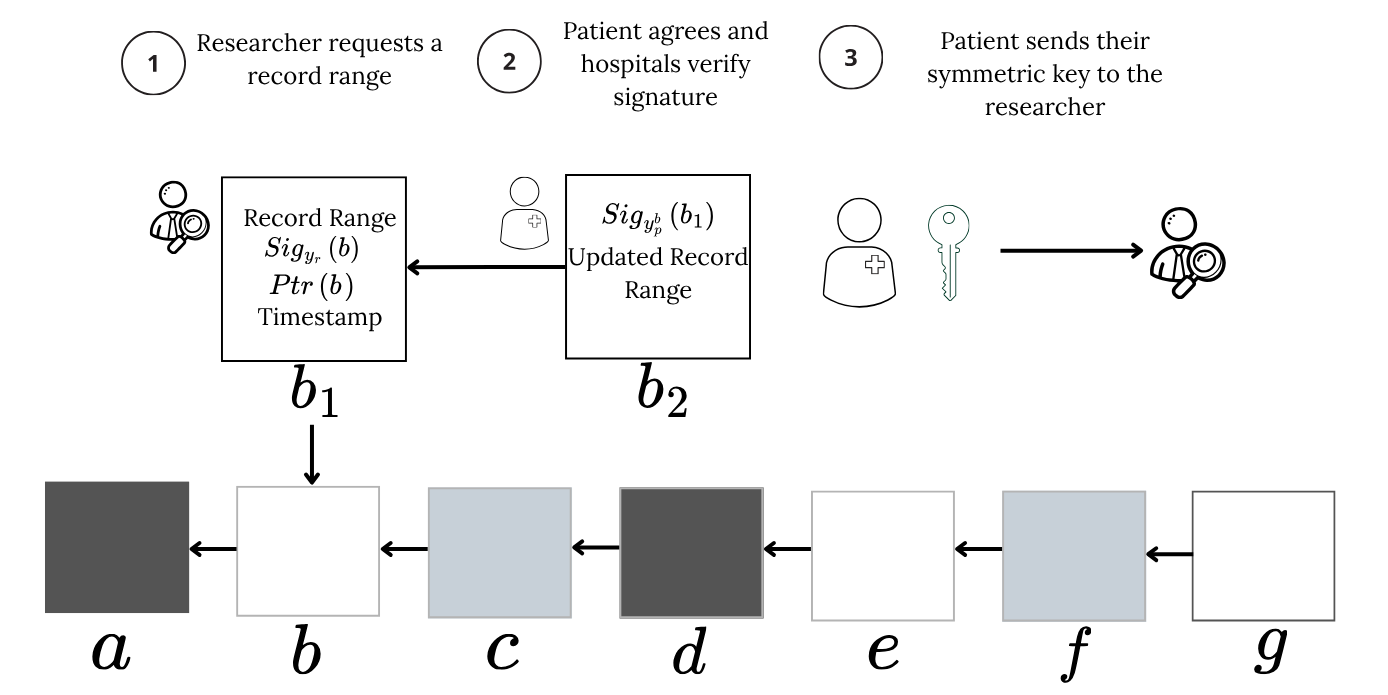}
    \caption{Process for Researcher Access}
    \label{researcherBlocks}
    
\end{figure*}

\subsubsection{PHR Block Confirmation}
In order to reach consensus, the miners in \Sys will need to verify that the information is accurate. The miners will confirm whether the ZKPs provided by the hospital and the patient in the header are valid. We use non-interactive Zero Knowledge Schnorr Proofs through the help of the Fiat-Shamir heuristic~\cite{fiat_shamir}. The miners will verify the ZKP transcripts on the block header. Once this has been done, the miners can verify that the signatures provided by the hospital and the patient are valid using the published public keys on the block. Once 50\% of the miners approve this information, they upload a new block onto \Sys.

\subsection{Phase II: Researcher Access}
After the blocks have been created and approved by the mining groups, researchers can comb through the blocks to find a patient that fits their specific area of interest by looking at the lifetime and temporary condition codes of a block. The full researcher process can be seen in Figure \ref{researcherBlocks}.

The researchers then find the block that fits their criteria by going through the patient's codes. From then on, the researcher can fork the block, which will act as an initial request. Within the initial fork, the researcher can request a time-based record range of PHRs found within the blockchain. It is important to note, however, that the researcher has no knowledge of what other blocks may or may not exist for that patient within \Sys. After requesting a specific range, the researcher includes the pointer to the block that they are forking from, $Ptr(b)$. They would then sign the block they forked with their public key, $Sig_{y_r}(b)$. All of this information gets stored on a new block, $b_1$, as a fork to the patient block. Once the researcher block has been created, it must be approved by the miners before it is uploaded to \Sys.

If the patient agrees to the researcher's request, they can create another block, $b_2$, where they sign $b_1$ with their public key, $Sig_{y^b_p}(b_1)$. The patient also writes the record range they wish to grant to the researcher in $b_2$. If they would like to narrow the researcher's range, they may do so.

Once the patient signs off on the block, they will give the researcher all the data they requested. They will need to give $3k + 2$ pieces of information, where $k$ is the number of blocks in the range. For each block $b$, they will need to give the researcher the symmetric key, $s^b_p$, a pointer to the data, $Ptr(D_b)$, and the hash of the data $H(D_b)$. For the first block in the range, block $b_f$, they will also need to give $H(\hat{b_f}-1)$, so that the researcher can calculate $H(\hat{b_f})$. Additionally, for the last block in the range, $b_l$, they will need to provide $n^{b_l}$, the nonce of $b_l$, to allow the researcher to compare their final hash with the hash on the blockchain. This also means the patient has to reveal $b_l$ in memory, however, $b_l$ is the only block that would need to be revealed. With only this information, the researcher can calculate $H(\hat{b})$ for each block in the range, and can verify that all the information is accurate using the final block. As we have implemented a continuous hashing formula, if even one of these values is tampered with, all subsequent hashes (including the final one stored on the blockchain with the nonce) will not match. This storage method prevents malicious researchers from going outside the allowed record range and makes it impossible to know which blocks they are actually getting information from. This also makes it impossible for the patient to omit any data as the hashes would not match, and it is unreasonable for a patient to compute a different input that hashes to the same output due to the security of the hashing algorithm~\cite{hashSecurity}.

\section{Experimental Evaluation}\label{sec:experimental_evaluation}
All experiments were performed locally on a computer cluster. We requested four tasks per node, two nodes, and 128gbs of memory.

\begin{figure*} [t]
    \centering
    \subfigure[2,000 Hospitals]{\includegraphics[width=0.32\textwidth]{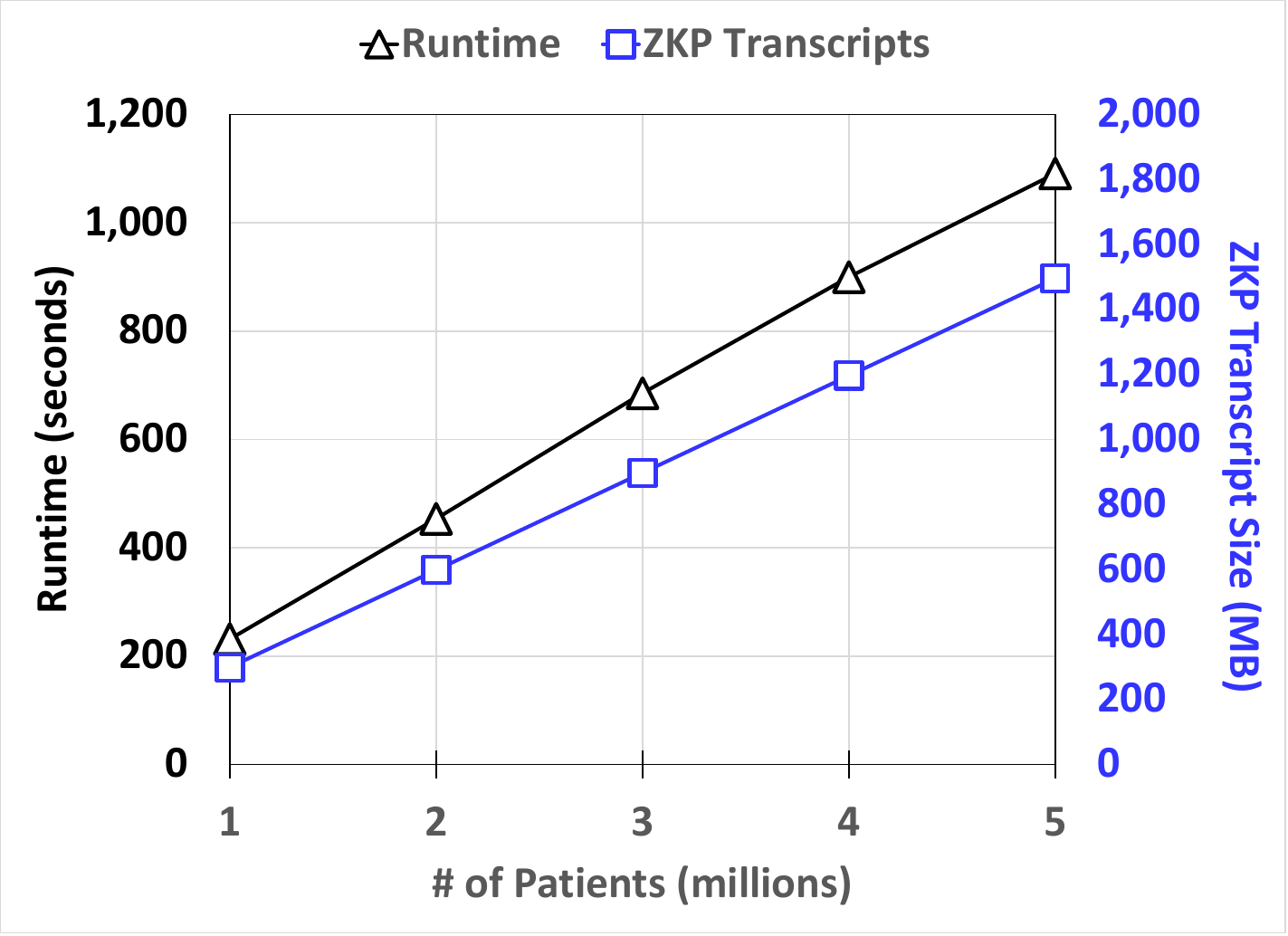}}
    \hfill
    \subfigure[4,000 Hospitals]{\includegraphics[width=0.32\textwidth]{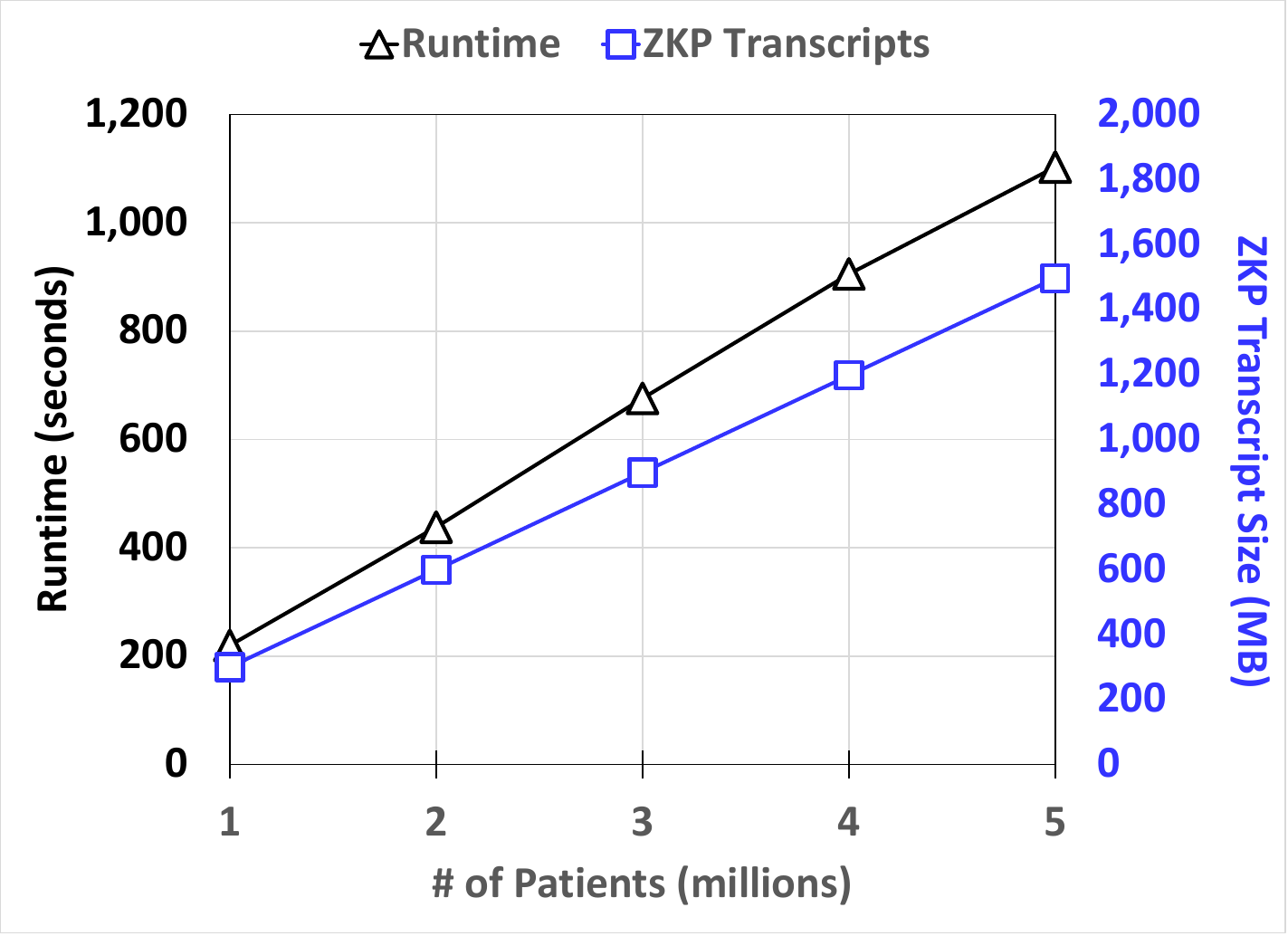}}
    \hfill
    \subfigure[6,000 Hospitals]{\includegraphics[width=0.32\textwidth]{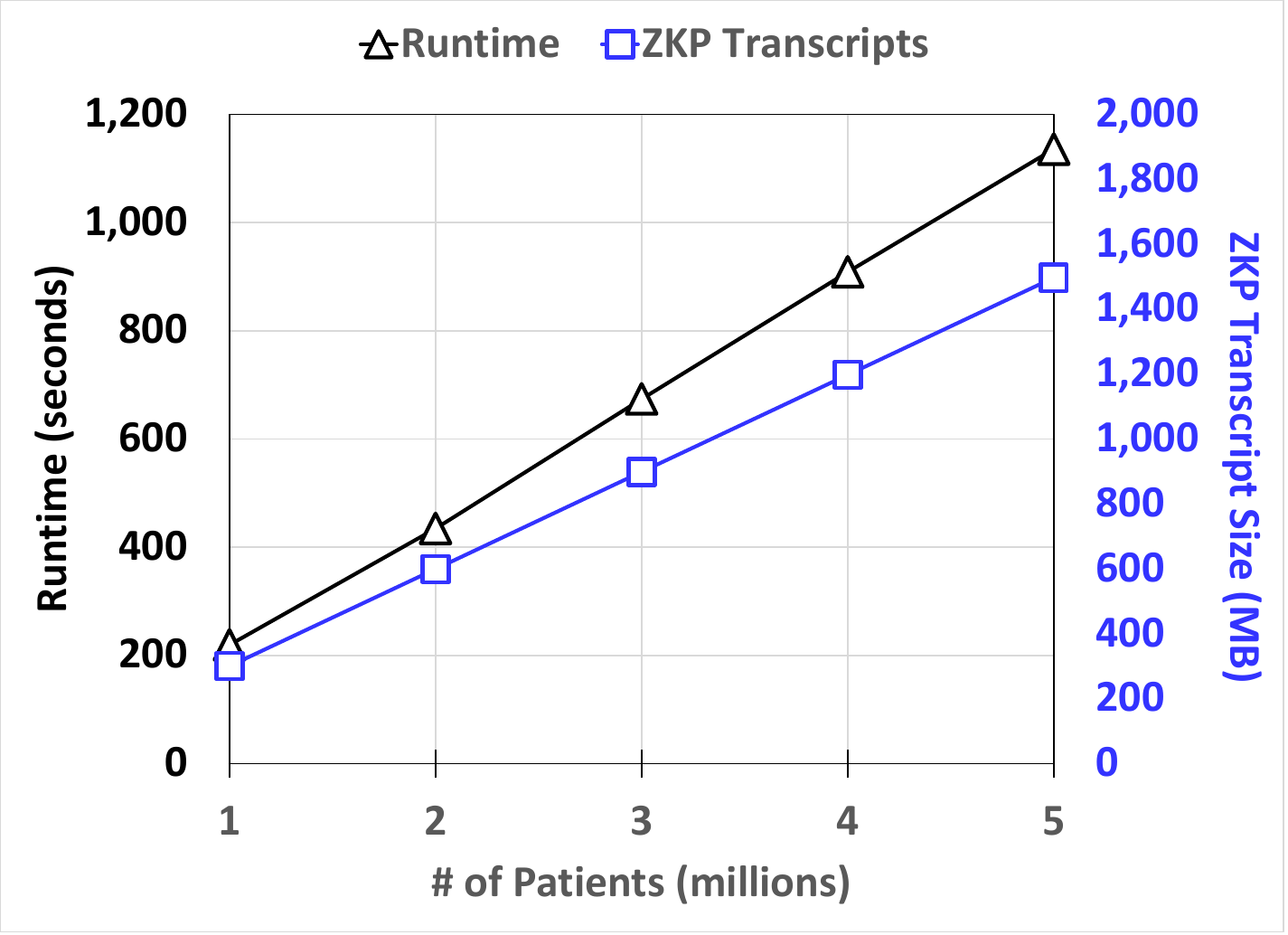}}
    \caption{Patient Block Creation}
    \label{patientBlockCreation}
\end{figure*}

\subsection{Patient Block Creation}
As \Sys gains more patients and hospitals in the registry, both the time and space required to create a new block will increase, potentially being infeasible for scalability. This experiment measures both the creation time of the block and the size of transcripts from a ZKP with respect to both the number of patients and the number of hospitals. 

We separated our data into three different graphs, shown on Figure~\ref{patientBlockCreation}. Each one contains the number of patients on the x-axis (from one to five million, incremented by one million), the runtime of the protocol on the primary y-axis (measured in seconds), and the size of the ZKP transcripts on the secondary y-axis (measured in megabytes). The first graph shows these ten data points for a set of two thousand hospitals, the second for four thousand, and the third for six thousand. 

As the bulk of the runtime is spent doing the ZKP, we would expect to see a runtime that is roughly $O(\#patients + \#hospitals)$. We observe in Figure~\ref{patientBlockCreation} that the runtime appears to match this prediction. Additionally, as the number of ZKPs being conducted is equal to the sum of number of patients and the number of hospitals, we would expect the ZKP transcript size to also grow at approximately $O(\#patients + \#hospitals)$. Looking at Figure~\ref{patientBlockCreation}, this observation appears to be reflected in the data, with the ZKP transcript size growing at appromately the same rate.



\subsection{Hospital Consensus Process}
Miners are essential to \Sys, where they are relied upon to verify the integrity of each block. The purpose of this experiment is to measure robustness by determining how long it would take for a block to be approved assuming that our mining group has been compromised. This experiment measures robustness with respect to the malicious percentage of miners, as well as the scalability with respect to the number of hospitals. We measures the amount of time it takes to reach consensus on a patient block in \Sys with a varying percentage of malicious miners, from 10\% to 40\%. Block approval happens when half of the miners accept the block. 

\begin{figure}[h]

    \centering
    \includegraphics[width = 0.9\linewidth]{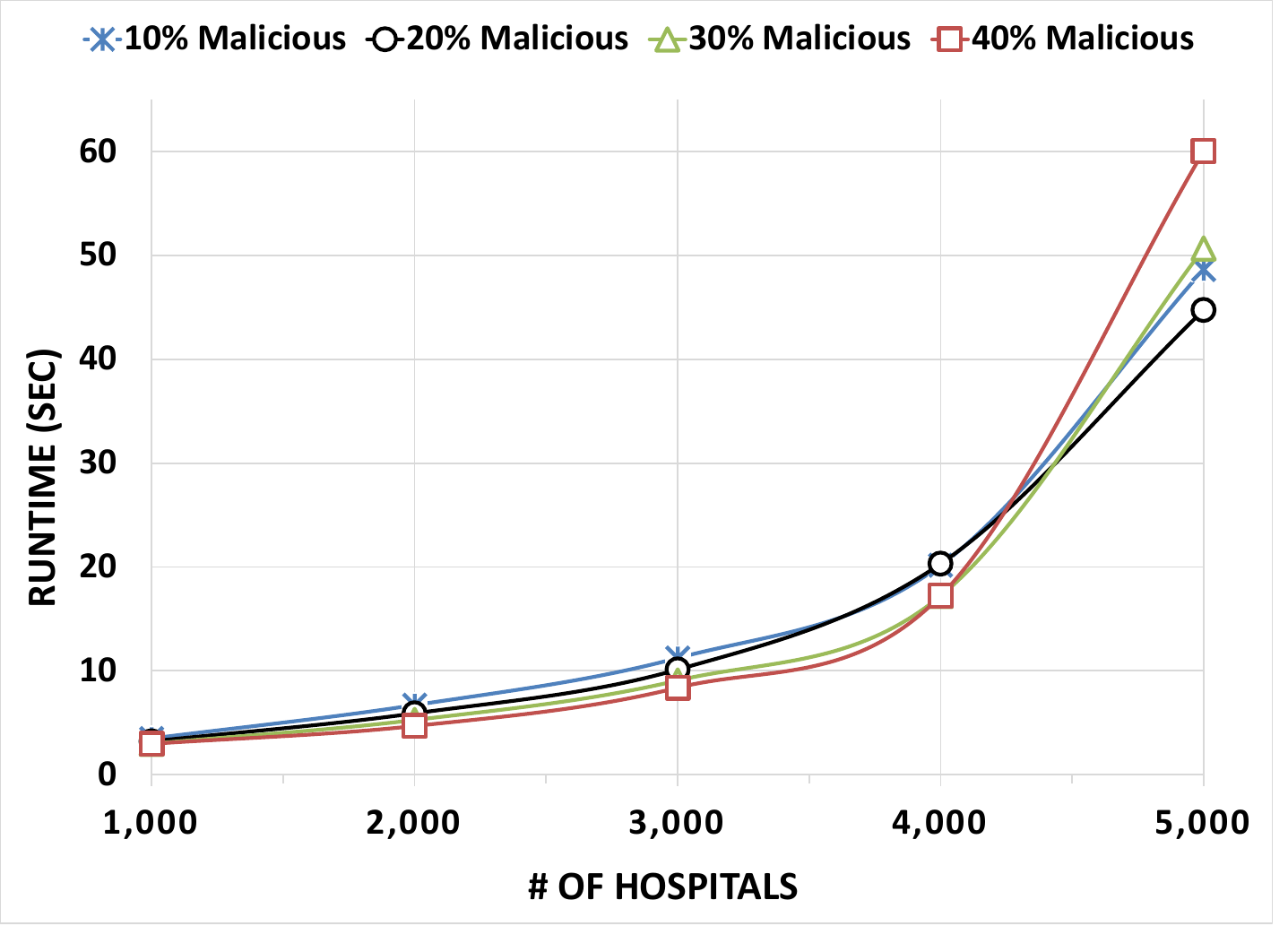}
    \caption{Hospital Consensus Process}
    \label{hospialConsensusProcess}
    
\end{figure}

Our results are shown in Figure~\ref{hospialConsensusProcess}. On the x-axis, we have the number of hospitals from one thousand to five thousand, incremented by one thousand each time. On the y-axis, we measured the runtime in seconds for ten, twenty, thirty, and forty percent of malicious hospitals. The number of folds for this experiment is four.


We observe in Figure~\ref{hospialConsensusProcess} that the runtime increases in $O(n^2)$ as the number of hospitals increases linearly, which is consistent with a model where every node propagates a signature to each other. We also observe that the runtime for different malicious levels with a fixed number of hospitals does not appear to significantly change.


\subsection{Researcher Access}
Researchers wanting to gain access to a patient's data have to go through a multi step process in order to be granted permission, including verification of requests by the miners. This experiment seeks to measure the robustness of the researcher access protocol with respect to malicious miners. Additionally, it seeks to measure the scalability of the protocol with respect to the percentage of malicious nodes. 

\begin{figure}[h]

    \centering
    \includegraphics[width = 0.9\linewidth]{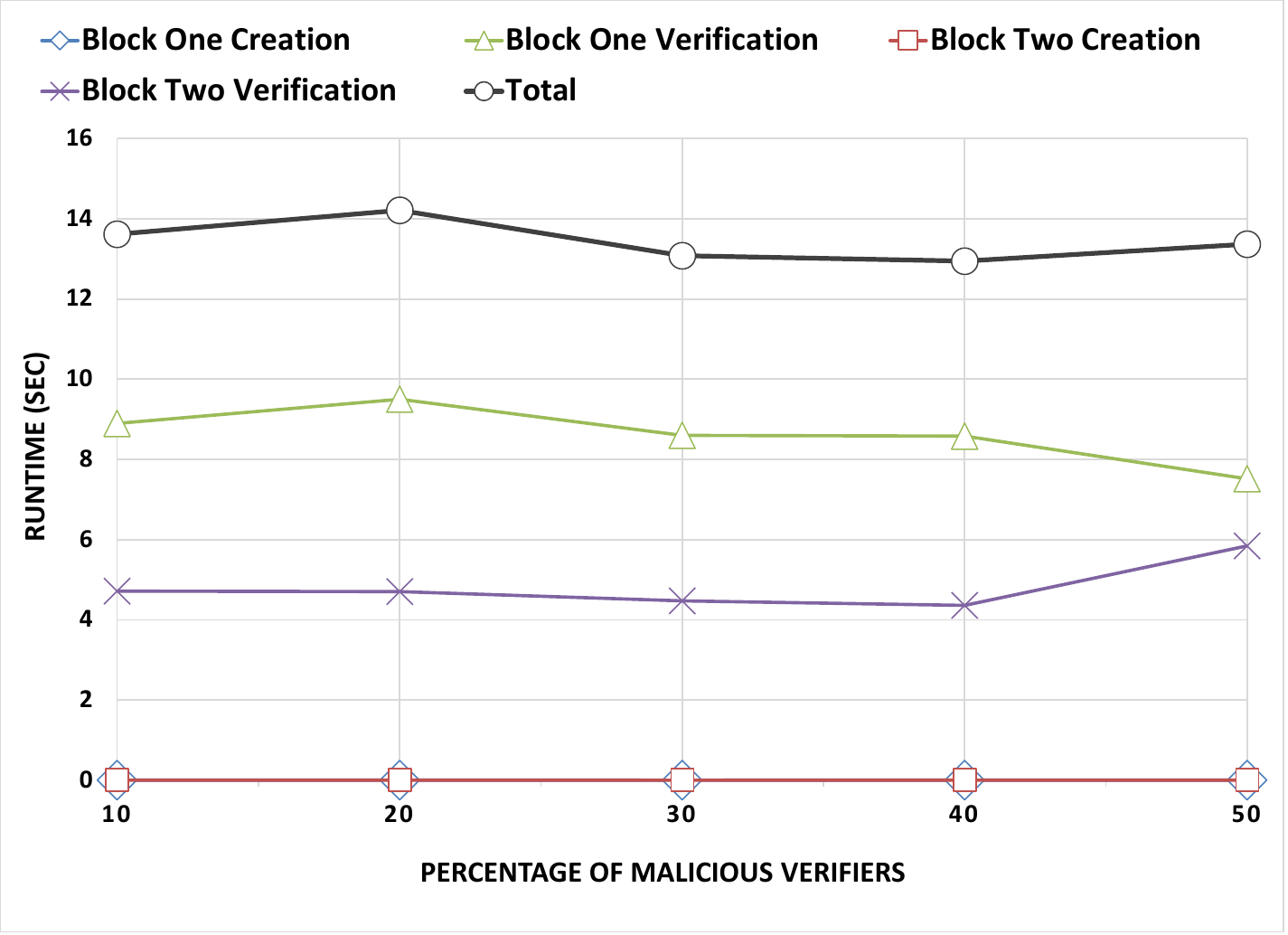}
    \caption{Researcher Access. Block One indicates the researcher request block, while Block Two indicates the patient approval block.}
    \label{fig:researcherAccessGraph}
    
\end{figure}

Our results for this experiment are located in Figure~\ref{fig:researcherAccessGraph}. On the x-axis, we measure the percentage of malicious miners, ranging from ten percent to fifty percent, incremented by ten percent. On the y-axis, we have the runtime of the whole operation in seconds. We fix the number of hospitals at six thousand for this experiment. We measured the runtime for four phases of the process, which include the creation and verification of the researcher request block (Block One), and the creation and verification of the patient approval block (Block Two). The number of folds for this experiment is four.


We observe that the creation time for both blocks is extremely quick, consistently less than $0.01$ seconds, and the creation times for both blocks are independent of the malicious percentage of the mining pool. Our verification runtime is relatively constant even with an increased malicious percentage. The reason for this is our assumption that malicious nodes immediately reject while honest nodes take their time to verify before approving. The total runtime of this process is approximately $O(1)$, showing that our system is robust for processing researcher requests with up to 50\% of malicious miners.


\section{Conclusion and Future Work}\label{sec:conclusion}

In this paper we proposed \Sys, a blockchain based system where patients can maintain total anonymity while providing PHR to researchers. \Sys implements many different privacy measures to protect patient anonymity. Participation is voluntary, and if a patient decides to enroll, they are given different public keys for each block, making it harder for a malicious user to connect different blocks to the same person. \Sys also stores only a hash of several different variables, rather than variable each separately, meaning a malicious user would need to know multiple correct variables to identify a block, including a single use nonce. As all the condition codes are made public on the blockchain, \Sys makes it easy for a researcher to identify blocks they may be interested in. By forking the desired block, they can request patients to share their PHR, and they can ensure the validity of the data provided using the hashes on the blockchain. \Sys is a robust system with respect to the amount of malicious miners up to fifty percent as demonstrated by the experiments we designed and ran. \Sys is also scalable with respect to the number of existing hospitals and patients in the registry. \Sys can have important consequences with managing access control and protecting patient privacy. \Sys has also been shown to be ethically virtuous when protecting the privacy of all users involved.

As a future work, it would be interesting to investigate how to support other types of queries besides just time-based range queries, such as count and/or conditional range queries. Another future work would be to investigate how a patient can satisfy a researcher query without revealing any block that belongs to that patient. Our proposed solution reveals one block if the range query does not require future blocks' information, and two if it does.



\section*{Acknowledgment}
This work is supported in part by the
National Science Foundation under award number 2349042.

\bibliographystyle{IEEEtranN}
\bibliography{Citations}   

\end{document}